\documentclass[conference]{IEEEtran}
\IEEEoverridecommandlockouts
\usepackage{bbm}
\usepackage{amsfonts}
\usepackage[dvips]{graphicx}
\usepackage{times}
\usepackage{cite}
\usepackage{amsmath}
\usepackage{array}
\usepackage{amssymb}

\newcommand{\bs}{\boldsymbol}

\usepackage{stfloats}
\usepackage{graphicx}
\usepackage{footnote}
\usepackage{booktabs}
\usepackage{array}
\usepackage{algorithm}
\usepackage{subeqnarray}
\usepackage{cases}
\usepackage{threeparttable}
\usepackage{color}
\usepackage{hyperref}
\usepackage{epstopdf}
\usepackage{algpseudocode}
\usepackage{bm}
\usepackage{multirow}
\usepackage[labelformat=simple]{subcaption}
\usepackage{adjustbox}

\usepackage{enumitem}

\makeatletter
\newcommand\fs@betterruled{%
  \def\@fs@cfont{\bfseries}\let\@fs@capt\floatc@ruled
  \def\@fs@pre{\vspace*{5pt}\hrule height.8pt depth0pt \kern2pt}%
  \def\@fs@post{\kern2pt\hrule\relax}%
  \def\@fs@mid{\kern2pt\hrule\kern2pt}%
  \let\@fs@iftopcapt\iftrue}
\floatstyle{betterruled}
\restylefloat{algorithm}
\makeatother

 \def\BibTeX{{\rm B\kern-.05em{\sc i\kern-.025em b}\kern-.08em T\kern-.1667em\lower.7ex\hbox{E}\kern-.125emX}}

\newlength{\bottomMargin}
\newtheorem{theorem}{\bf Theorem}

\newtheorem{lemma}{\bf Lemma}
\newtheorem{definition}{\bf Definition}

\allowdisplaybreaks

\begin{document}

\title{On the Age of Information in Internet of Things Systems with Correlated Devices}

\author{\IEEEauthorblockN{Bo Zhou and Walid Saad}
\IEEEauthorblockA{Wireless@VT, Bradley Department of Electrical and Computer Engineering, Virginia Tech, Blacksburg, VA, USA,\\
Emails: \{ecebo, walids\}@vt.edu
}
\\
 \thanks{This research was supported by the  U.S. Office of Naval Research (ONR) under Grant N00014-15-1-2709.}
 \\[-0.0ex]
}

\maketitle

\begin{abstract}
In this paper, a real-time Internet of Things (IoT) monitoring system is considered in which multiple IoT devices must transmit timely updates on the status information of a common underlying physical process to a common destination. In particular, a real-world IoT scenario is considered in which multiple (partially) observed status information by different IoT devices are required at the destination, so that the real-time status of the physical process can be properly re-constructed. By taking into account such correlated status information at the IoT devices, the problem of IoT device scheduling is studied in order to jointly minimize the average age of information (AoI) at the destination and the average energy cost at the IoT devices. Particularly, two types of IoT devices are considered: Type-I devices whose status updates randomly arrive and type-II devices whose status updates can be generated-at-will with an associated sampling cost. This stochastic problem is formulated as an infinite horizon average cost Markov decision process (MDP). The optimal scheduling policy is shown to be threshold-based with respect to the AoI at the destination, and the threshold is non-increasing with the channel condition of each device. For a special case in which all devices are type-II, the original MDP can be reduced to an MDP with much smaller state and action spaces. The optimal policy is further shown to have a similar threshold-based structure and the threshold is non-decreasing with an energy cost function of the devices. Simulation results illustrate the structure of the optimal policy and show the effectiveness of the optimal policy compared with a myopic baseline policy.
\end{abstract}

 \section{Introduction}
 In emerging Internet of Things (IoT) applications \cite{8533634,wang2013intelligent,7875131}, such as environmental monitoring, drone control, and smart surveillance, it is critical to ensure a timely delivery of status information of the underlying physical processes that are being monitored by various of IoT devices.
However, conventional performance measures, such as delay and throughput, cannot properly capture the timeliness of the status information from the perspective of the information destination.
In particularly, minimizing the delay or maximizing the throughput, may not maintain the status information at the destination as fresh as possible\cite{8000687}.

Recently, the notion of \emph{age of information} (AoI) has been proposed as a promising candidate for quantifying the freshness of IoT status information\cite{6195689}.
Typically, the AoI is defined as the time elapsed since the most recently received status packet at the destination was collected by the IoT devices, and hence it naturally captures how fresh the information is from the destination’s perspective.
Due to the importance of information freshness in the IoT, the concept of AoI has
received significant attention in a variety of scenarios that include
 multi-user networks\cite{8469047,maatouk2019minimizing,8919867,ceran2018reinforcement}, multi-hop networks \cite{8732378,8262777}, IoT monitoring systems \cite{8778671,8938128,8894836}, energy harvesting systems\cite{8822722,abd2019reinforcement,8437496}, cognitive networks \cite{wang2020minimizing,globasip19}, and remote estimation systems \cite{ornee2019sampling}.

These existing works, e.g.,\cite{8000687,6195689,8469047,maatouk2019minimizing,8919867,ceran2018reinforcement,8732378,8262777,8778671,8938128,8822722,abd2019reinforcement,8437496,wang2020minimizing,globasip19,ornee2019sampling,8894836}, assume that different devices are associated with different independent physical processes and, hence, the AoI at the destination will change upon receiving a ``fresher'' status update from only one device.
However, for certain IoT applications, updates from different devices could be correlated in the sense that multiple devices can be associated with one common physical process. In these scenarios, the (partially) observed status updates  by multiple devices are required at the destination, so that the real-time status information can be properly re-constructed.
For example, for smart camera networks\cite{wang2013intelligent}, in which multiple cameras with possible overlapping fields of view are monitoring a common scene, the remote monitor can only re-construct the full image of the scene after receiving the images acquired from multiple cameras.
Thus, for such scenarios, the AoI at the destination will only change upon receiving the fresher status updates from multiple devices, which renders the existing approaches in \cite{8000687,6195689,8469047,maatouk2019minimizing,8919867,ceran2018reinforcement,8732378,8262777,8778671,8938128,8822722,abd2019reinforcement,8437496,wang2020minimizing,globasip19,ornee2019sampling,8894836} no longer applicable.

Recently, the works in \cite{8849480} and \cite{8463500} considered remote estimation systems with spatially correlated processes and proposed updating strategies to improve the estimation error.
For monitoring systems with common observations, the work in \cite{8723545} proposed scheduling policies to minimize the average AoI and the work in \cite{javani2019age} analyzed the average AoI from a queueing theoretic perspective. However, the authors in \cite{8723545} and \cite{javani2019age} assume that the AoI at the destination changes if a fresh status update from only one device is delivered.
In \cite{8825510}, the authors study joint node assignment and scheduling to minimize the maximum peak AoI for wireless camera networks with  multi-view processing. However, the status packets at each camera are assumed be already enqueued in its buffer.
 Clearly, it remains unknown how to schedule the IoT devices to minimize the average AoI for IoT monitoring systems, in which multiple updates from different devices are required at the destination and the status packets can randomly arrive or be generated at the devices.

The main contribution of this paper is, thus, a novel design of the optimal IoT device scheduling policy that jointly minimizes the average AoI at the destination and the energy cost at the IoT devices for a real-time IoT monitoring system with correlated devices.
We consider two types of correlated devices: type-I devices whose status updates randomly arrive and type-II devices whose status updates can be generated-at-will.
We formulate this problem as an infinite horizon average cost Markov decision process (MDP).  By exploiting the properties of the AoI dynamics and analyzing the monotonicity property of the value function for the MDP, we show that the optimal scheduling policy is threshold-based with respect to the AoI at the destination and the threshold is  non-increasing with the channel condition of each device.
For a special case with only type-II devices, we reduce the original MDP to an MDP with much smaller state and action spaces, and we show that the optimal scheduling policy is also threshold-based and the threshold is non-decreasing with an energy cost function of the IoT devices.
Simulation results validate the structural analysis of the optimal policy and show its effectiveness over a myopic baseline policy.

The rest of this paper is organized as follows. Section II introduces the system model. In Section III, we present the problem formulation and the optimality equation. In Section IV, we characterize the structural properties of the optimal policy. Section V presents the analysis for a special IoT case.
Simulation results and analysis are provided in Section VI. Finally, conclusions are drawn in Section VII.

\section{System Model}
\begin{figure}[!t]
\begin{centering}
\includegraphics[scale=.25]{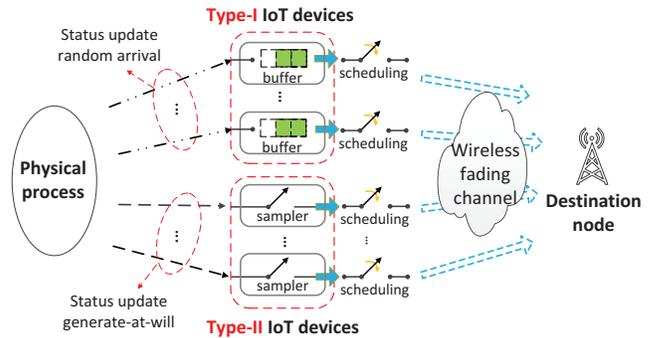}
 \caption{Illustration of a real-time monitoring system with two types of correlated IoT devices.}\label{fig:system}
\end{centering}
\end{figure}

 As illustrated in Fig.~\ref{fig:system}, we consider a real-time IoT monitoring system in which a set $\mathcal{N}$ of $N$ IoT devices can collect the real-time status information of a common  underlying physical process and update the status packets to a common remote destination node (e.g., a base station or fusion center).
In our model, each IoT device can only observe partial status information of its underlying physical process, and, hence, multiple status packets from different IoT devices are required at the destination so that the real-time status of the physical process can be properly re-constructed.
Let $M\geq 2$ be the number of different IoT devices that are needed at the destination for successful status reconstruction.
Such a scenario with \emph{correlated} IoT devices is relevant to practical IoT applications, for example, smart camera networks with overlapping fields of view\cite{wang2013intelligent}.
Note that, this is different from prior works \cite{8000687,6195689,8469047,maatouk2019minimizing,8919867,ceran2018reinforcement,8732378,8262777,8778671,8938128,8822722,abd2019reinforcement,8437496,wang2020minimizing,globasip19,ornee2019sampling,8894836} in which different  devices are associated with different \emph{independent} physical processes and the status packet from only one device is needed to re-construct the real-time status of the corresponding physical process.

We consider a discrete-time system with unit slot length that is indexed by $t=1,2,\cdots$.
For each IoT device $n\in\mathcal{N}$, let $h_n(t)\in\mathcal{H}_n$ be the channel state at time slot $t$, where $\mathcal{H}_n$ is the finite channel state space of device $n$.
We assume i.i.d. block fading wireless channels for all devices, and we consider the channel state processes $\{h_n(t)\}(n\in\mathcal{N})$ to be mutually independent.
Let $\bs{h}(t)\triangleq(h_n(t))_{n\in\mathcal{N}}\in\mathcal{H}\triangleq\prod_{n\in\mathcal{N}}\mathcal{H}_n$ be the system channel state vector at slot $t$, where $\mathcal{H}$ is the system channel state space.

We consider two types of IoT devices, namely, a set $\mathcal{N}_1\subseteq\mathcal{N}$ of $N_1$ type-I devices and a set $\mathcal{N}_2=\mathcal{N}\setminus\mathcal{N}_1$ of $N_2$ type II devices.
For each type-I device, at the beginning of each slot, a (partially observed) status packet (if any) of the underlying physical process randomly arrives and will be stored in its buffer while replacing the older one (if any) that was not transmitted.
We assume that the status packet arrivals for different type-I devices are mutually independent, and for each type-I device $n\in\mathcal{N}_1$, the process of the status packet arrivals is an i.i.d. Bernoulli process with mean rate $\lambda_n\in[0,1]$.
On the other hand, at each slot, each type-II device can collect the partially observed real-time status of the physical process and generate a status packet at will.
We assume that the time for generating a status packet is negligible for each type-II device, as is commonly used in the literature (e.g., \cite{8000687,8919867,ceran2018reinforcement} and \cite{8262777}).

\subsection{Monitoring Model}
In each slot, the network must decide which IoT devices must be scheduled so as to update their status packets to the destination.
For each IoT device $n\in\mathcal{N}$, let $u_n(t)\in\{0,1\}$ be the scheduling action at slot $t$, where $u_n(t)=1$ indicates device $n$ is scheduled to transmit its status packet and $u_n(t)=0$, otherwise.
In each slot, we consider that at most $M$ IoT devices can transmit their status packets concurrently without transmission collisions over different orthogonal channels.
Thus, we have $\sum_{n\in\mathcal{N}}u_n(t)\leq M$ for all $t$.
Let $\bs{u}(t)\triangleq (u_n(t))_{n\in\mathcal{N}}\in\mathcal{U}$ be the system scheduling action at slot $t$, where $\mathcal{U}\triangleq \{(u_n)_{n\in\mathcal{N}} | u_n\in\{0,1\},~\forall n\in\mathcal{N} \text{~and~} \sum_{n\in\mathcal{N}} u_n\leq M\}$ is the feasible system scheduling action space.

For each device $n\in\mathcal{N}$, let $C_n^u(h_n)$ be the minimum transmission power required by device $n$ to deliver its status packet to the destination within a slot when its channel state is $h_n$. Without loss of generality, we assume that $C_n^u(h_n)$ is non-increasing with $h_n$, i.e., a large $h_n$ implies better channel condition.
Each type-II device $n\in\mathcal{N}_2$ will incur an non-zero sampling cost $C_n^s$ for generating a status packet.
Note that, there is no such sampling cost for any type-I devices. For notational convenience, we set $C_n^s=0$ for each type-I device $n\in\mathcal{N}_1$.
Then, for each device $n\in\mathcal{N}$, the energy cost associated with action $u_n(t)$ under channel state $h_n(t)$ is given by: $C_n(u_n(t),h_n(t))=u_n(t)(C_n^s + C_n^u(h_n(t)))$.

\subsection{Age of Information Model}
The AoI is adopted as the performance metric to quantify the freshness of the status information update  at the destination, which is defined as the time elapsed since the most recent status update of the physical process was received (re-constructed) at the destination.
Let $\Delta(t)$ be the AoI at the destination at the beginning of slot $t$.
Note that, a type-I device can only transmit its currently available status packet to the destination, and, thus, the AoI at the destination $\Delta(t)$ would depend on the age of the status packet in the buffer of each type-I device.
For each type-I device $n\in\mathcal{N}_1$, let $A_n(t)$ be the AoI at device $n$ at the beginning of slot $t$.
We denote by $\hat{A}_n$ and $\hat{\Delta}$ the upper limits of the AoI at type-I device $n$ and the AoI at the destination, respectively.
We assume that $\hat{A}_n$ and $\hat{\Delta}$ are finite, as a status packet with an infinite age is not meaningful for a real-time monitoring system.
Let $\mathcal{A}\triangleq\prod_{n\in\mathcal{N}_1}\mathcal{A}_n$ and $\varDelta\triangleq\{1,2,\cdots,\hat{\Delta}\}$ be, respectively, the state spaces for the AoI at type-I devices and the AoI at the destination, where $\mathcal{A}_n\triangleq\{1,2,\cdots,\hat{A}_n\}$.
Let $\bs{A}(t)\triangleq(A_n(t))_{n\in\mathcal{N}_1}\in\mathcal{A}$ be the system AoI state at type-I devices.
Note that, a type-II device can immediately generate a status packet if scheduled.
Thus, with a slight abuse of notation, we also denote by $A_n(t)$  the AoI at each type-II device $n\in\mathcal{N}_2$, and we set $A_n(t)=0$ for all $n\in\mathcal{N}_2$ and $t$.

We can now show how $\bs{A}(t)$ and $\Delta(t)$ evolve with the scheduling action $\bs{u}(t)$. For the AoI at a type-I device, if there is a status packet arriving at device $n\in\mathcal{N}_1$ at slot $t$, then the AoI $A_n(t)$ will decrease to one, otherwise, the AoI will increase by one. Thus, the dynamics of  $A_n(t)$ for each type-I device $n\in\mathcal{N}_1$ are given by:
\begin{align}\label{eqn:aoi_device}
A_n(t+1)=\begin{cases} 1,  \hspace{2mm}\text{if a status update arrives at device $n$ at $t$},\\
                \min\{A_n(t)+1,\hat{A}_n\}, \text{~otherwise.}
      \end{cases}
\end{align}

For the AoI at the destination, if fewer than $M$ devices are scheduled at slot $t$, then the AoI $\Delta(t)$ will increase by one, otherwise, to re-construct the status information of the physical process according to the received status packets from $M$ devices, the AoI  will be the largest age of the received status packets from all the scheduled devices plus one. Thus, the AoI dynamics of the destination are given by:
\begin{align}\label{eqn:aoi_destination}
 &\Delta(t+1)\\
 &=\begin{cases} \min\{\max\limits_{n\in\mathcal{N}} u_n(t)A_n(t)+1,\hat{\Delta}\},&~\text{if}~\sum_{n\in\mathcal{N}}{u_n}(t)= M,\\
                \min\{\Delta(t)+1,\hat{\Delta}\},&~\text{otherwise.}
      \end{cases}\nonumber
\end{align}

Clearly, it is not optimal to choose the scheduling action $\bs{u}$ such that $0<\sum_{n\in\mathcal{N}}{u_n}< M$, as the destination will not benefit from receiving status packets from fewer than $M$ devices, yielding energy waste at the devices.
Note that, the dynamics in \eqref{eqn:aoi_destination} are different from those in prior art \cite{8000687,6195689,8469047,maatouk2019minimizing,8919867,ceran2018reinforcement,8732378,8262777,8778671,8938128,8822722,abd2019reinforcement,8437496,wang2020minimizing,globasip19,ornee2019sampling,8894836} in which the status packet from only one device is required for the destination to re-construct the real-time status of the physical process.

\section{Problem Formulation and Optimality}
\subsection{Problem Formulation}
We aim at controlling the IoT devices scheduling process so as to jointly minimize the average AoI at the destination and the energy cost at the IoT devices, under correlated IoT devices.
Given an observed system AoI state at type-I devices $\bs{A}$, the AoI state at the destination $\Delta$, and the system channel state $\bs{h}$, we can determine the system scheduling action  $\bs{u}$ according to the following policy.
\begin{definition}\label{definition:stationary_policy_IoT}
A \emph{feasible stationary scheduling policy} $\pi$ is  a mapping from the state spaces for the AoI at type-I devices, the AoI at the destination, and the system channel $\mathcal{A}\times\varDelta\times\mathcal{H}$ to the feasible system scheduling action space $\mathcal{U}$, i.e., $\pi(\bs{A},\Delta,\bs{h})=\bs{u}$.
\end{definition}

Under the dynamics in \eqref{eqn:aoi_device} and \eqref{eqn:aoi_destination}, the induced random process $\{\bs{A}(t),\Delta(t),\bs{h}(t)\}$ for a given feasible stationary scheduling policy $\pi$ is a controlled Markov chain with the following transition probability:
\begin{align}\label{eqn:trans_prob}
&\Pr[\bs{A}',\Delta',\bs{h}'|\bs{A},\Delta,\bs{h},\bs{u}]\nonumber\\
&=\prod_{n\in\mathcal{N}_1}\Pr[A'_n| A_n]\Pr[\Delta'|\bs{A},\Delta,\bs{u}]\prod_{n\in\mathcal{N}}\Pr[h'_n],
\end{align}
where
\begin{align}
&\Pr[A'_n| A_n]
=\Pr[A_n(t+1)=A_n| A_n(t)=A_n]\nonumber\\
&=\begin{cases}
\lambda_n,  &~\text{if}~A_n' = 1,\\
1-\lambda_n, &~\text{if}~A_n' = \min\{A_n+1,\hat{A}_n\},\\
                0, &~\text{otherwise,}
      \end{cases}
\end{align}
and
\begin{align}
&\Pr[\Delta'|\bs{A},\Delta,\bs{u}]\nonumber\\
&=\Pr[\Delta'(t+1)=\Delta'|\bs{A}(t+1)=\bs{A},\Delta(t)=\Delta,\bs{u}(t)=\bs{u}]\nonumber\\
&=\begin{cases}
1,  &~\text{if}~ \Delta'\text{satisfies~the~dynamics of~in~\eqref{eqn:aoi_destination}},\\
                0, &~\text{otherwise.}
      \end{cases}
\end{align}

Then, under a feasible stationary policy $\pi$ and a given initial system state $(\bs{A}_1,\Delta_1,\bs{h}_1)\in\mathcal{A}\times\varDelta\times\mathcal{H}$, the average AoI at the destination and the average energy cost for device $n\in\mathcal{N}$ are respectively given by:
\begin{align*}
&\bar{\Delta}^{\pi}\triangleq\limsup_{T\to\infty}\frac{1}{T}\sum_{t=1}^{T}  \mathbb{E} \left[\Delta(t)| \bs{A}_1,\Delta_1,\bs{h}_1\right],\\
&\bar{C}_n^{\pi}\triangleq\limsup_{T\to\infty}\frac{1}{T}\sum_{t=1}^{T}  \mathbb{E} \left[u_n(t)(C^s_n+C^u_n(h_n(t))| \bs{A}_1,\Delta_1,\bs{h}_1\right],
\end{align*}
where the expectations are taken with respect to the measure induced by policy $\pi$.
Our goal is to find the optimal scheduling policy $\pi$ that jointly minimizes the average AoI at the destination and the average energy cost for all IoT devices. By using the widely used weighted-sum method for multi-objective optimization problems \cite{deb2014multi}, we formulate the following problem:
\begin{align}
\min_{\pi}\bar{\Delta}^{\pi} + \sum_{n\in\mathcal{N}}\beta_n \bar{C}_n^{\pi},\label{eqn:mdp}
\end{align}
where $\pi$ is a feasible stationary policy in Definition~\ref{definition:stationary_policy_IoT} and $\beta_n>0$ for all $n\in\mathcal{N}$ are the weighting factors on the average energy cost for device $n$, mimicking the soft constraints on the average energy cost for each device.
The problem in \eqref{eqn:mdp} is an infinite horizon average cost MDP. To guarantee the existence of optimal stationary polices, we focus on stationary unichain polices, as is commonly done (e.g.,\cite{ceran2018reinforcement} and \cite{djonin2007mimo}).

\subsection{Optimality Equation}
According to \cite[Propositions 5.2.1, 5.2.3, and 5.2.5]{bertsekas4}, the optimal scheduling policy $\pi^*$ can be obtained by solving the following Bellman equation:
\begin{align}\label{eqn:bellman}
&  \theta + V(\bs{A},\Delta,\bs{h}) = \min_{\bs{u}\in\mathcal{U}} \Big\{\Delta + \sum_{n\in\mathcal{N}}\beta_nu_n(C^s_n+C^u_n(h_n))  \nonumber\\
&\hspace{3mm} + \sum_{\bs{A}',\Delta,\bs{h}'} \Pr[\bs{A}',\Delta',\bs{h}'| \bs{A},\Delta,\bs{h},\bs{u}] V(\bs{A}',\Delta',\bs{h}')\Big\},
\end{align}
 where $\Pr[\bs{A}',\Delta',\bs{h}'| \bs{A},\Delta,\bs{h},\bs{u}]$ is given by \eqref{eqn:trans_prob},  $\theta$ is the optimal value for all initial system states $(\bs{A}_1,\Delta_1,\bs{h}_1)\in\mathcal{A}\times\varDelta\times\mathcal{H}$, and $V(\bs{A},\Delta,\bs{h})$ is the value function.
 The optimal policy achieving the optimal value $\theta$ is given by:
\begin{align}\label{eqn:optimal-policy}
&  \pi^*(\bs{A},\Delta,\bs{h}) = \arg\min_{\bs{u}\in\mathcal{U}} \Big\{\Delta + \sum_{n\in\mathcal{N}}\beta_nu_n(C^s_n+C^u_n(h_n))  \nonumber\\
&\hspace{1mm} + \sum_{\bs{A}',\Delta,\bs{h}'} \Pr[\bs{A}',\Delta',\bs{h}'| \bs{A},\Delta,\bs{h},\bs{u}] V(\bs{A}',\Delta',\bs{h}')\Big\}.
\end{align}

It can be seen that obtaining the optimal policy $\pi^*$ in \eqref{eqn:optimal-policy} requires determining the value function $V(\cdot)$ by solving the Bellman equation in \eqref{eqn:bellman}, for which, there is generally no closed-form solution. Moreover, the numerical solutions (e.g., value iteration) do not usually provide many design insights, which further motivate us to study the structural properties of the optimal policy $\pi^*$ for a better understanding of the system.

\section{Structural Properties of the Optimal Policy}\label{sec:structure}
We characterize the structural properties of the optimal policy $\pi^*$ to the MDP in \eqref{eqn:mdp}.
First, by using the dynamics in \eqref{eqn:aoi_device} and \eqref{eqn:aoi_destination}, we first show the monotonicity property of the value function $V(\bs{A},\Delta,\bs{h})$ in the following Lemma.\footnote{All proof are omitted due to space limitations.}
\begin{lemma}\label{lemma:propertyV}
$V(\bs{A},\Delta,\bs{h})$ is non-decreasing with $A_n$ for all $n\in\mathcal{N}_1$ and $\Delta$, and is non-increasing with $h_n$ for all $n\in\mathcal{N}$.
\end{lemma}

Then, we present the structure of optimal policy $\pi^*$. Before that, we define the following function:
\begin{align}\label{eqn:threshold}
\phi_{\bs{u}}(\bs{A},\bs{h})\triangleq\begin{cases}\min\Phi_{\bs{u}}(\bs{A},\bs{h}),  & \text{if}~\Phi_{\bs{u}}(\bs{A},\bs{h})\neq\emptyset, \\
            +\infty,  &\text{otherwise},
  \end{cases},
\end{align}
where $\Phi_{\bs{u}}(\bs{A},\bs{h})\triangleq\{\Delta| \Delta\in\varDelta~\text{and}~J(\bs{A},\Delta,\bs{h},\bs{u})\leq J(\bs{A},\Delta,\bs{h},\bs{v}),~\forall\allowbreak \bs{v}\in\mathcal{U}\text{~and~}\bs{v}\neq \bs{u}\}$ and $J(\bs{A},\Delta,\bs{h},\bs{u})\allowbreak \triangleq \allowbreak\Delta + \sum_{n\in\mathcal{N}}\beta_nu_n(C^s_n+C^u_n(h_n))
+ \sum_{\bs{A}',\Delta,\bs{h}'} \Pr[\bs{A}',\Delta',\bs{h}'| \bs{A},\Delta,\bs{h},\bs{u}] V(\bs{A}',\Delta',\bs{h}')$, which is related to the right-hand side of the Bellman equation in \eqref{eqn:bellman}.

\begin{theorem}\label{theorem:optimal}
If $\exists n\in\mathcal{N}$, such that $u_n^*=1$, then $\pi^*(\bs{A},\Delta,\bs{h})=\bs{u}^*$ for all $(\bs{A},\Delta,\bs{h})\in\mathcal{A}\times\varDelta\times\mathcal{H}$ such that
\begin{equation}
\Delta\geq \phi_{\bs{u}^*}(\bs{A},\bs{h}).
\end{equation}
Moreover, $\phi_{\bs{u}^*}(\bs{A},\bs{h})$ is non-increasing with $h_n$, $\forall n\in\mathcal{N}$.
\end{theorem}

\begin{figure}[!t]
\begin{minipage}[h]{0.45\linewidth}
\centering
       \includegraphics[scale=0.26]{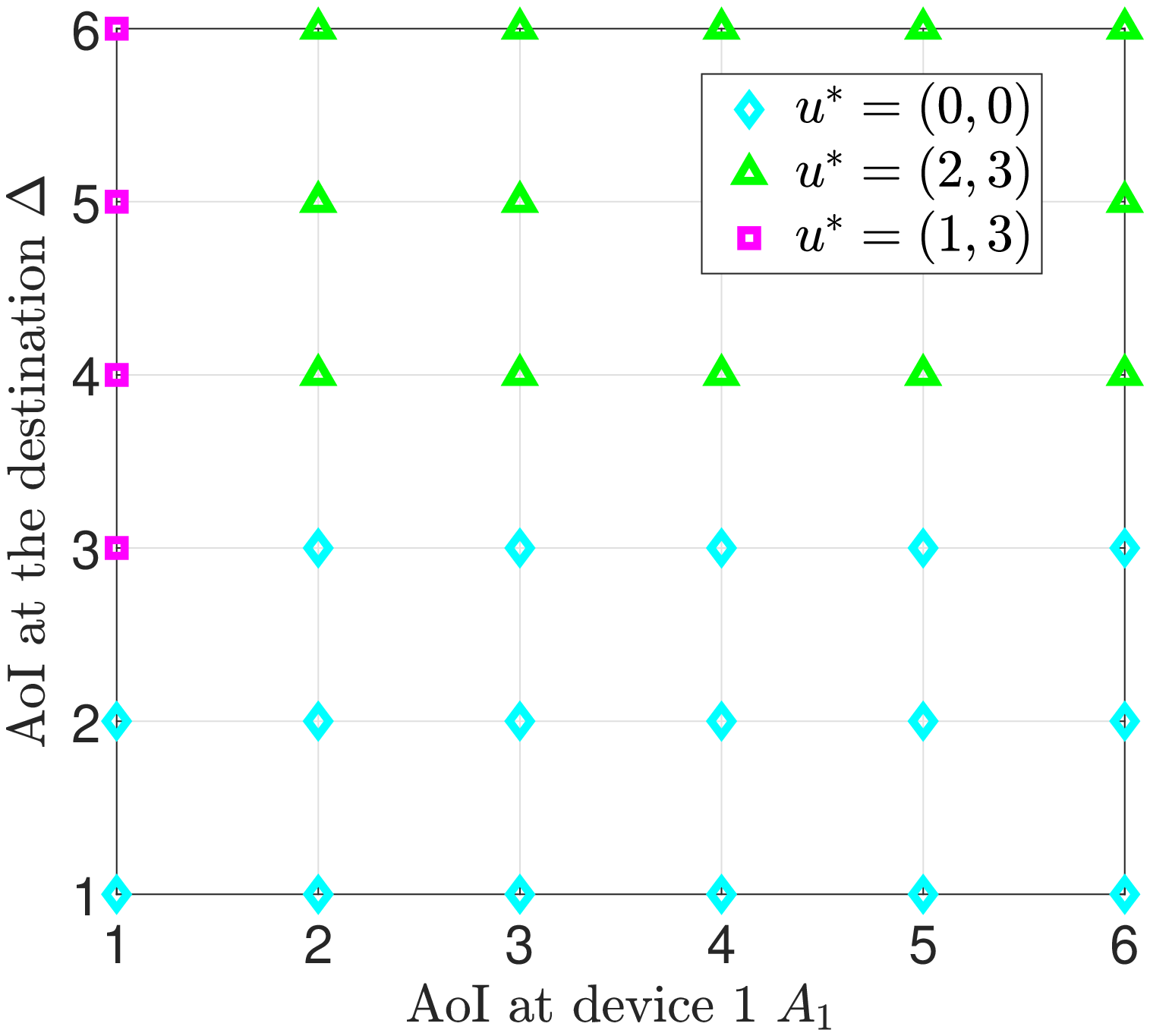}
\subcaption{Fixed $\bs{A}\setminus A_1$ and $\bs{h}$.}\label{fig:structure-A1}
\end{minipage}
\begin{minipage}[h]{0.45\linewidth}
\centering
        \includegraphics[scale=0.26]{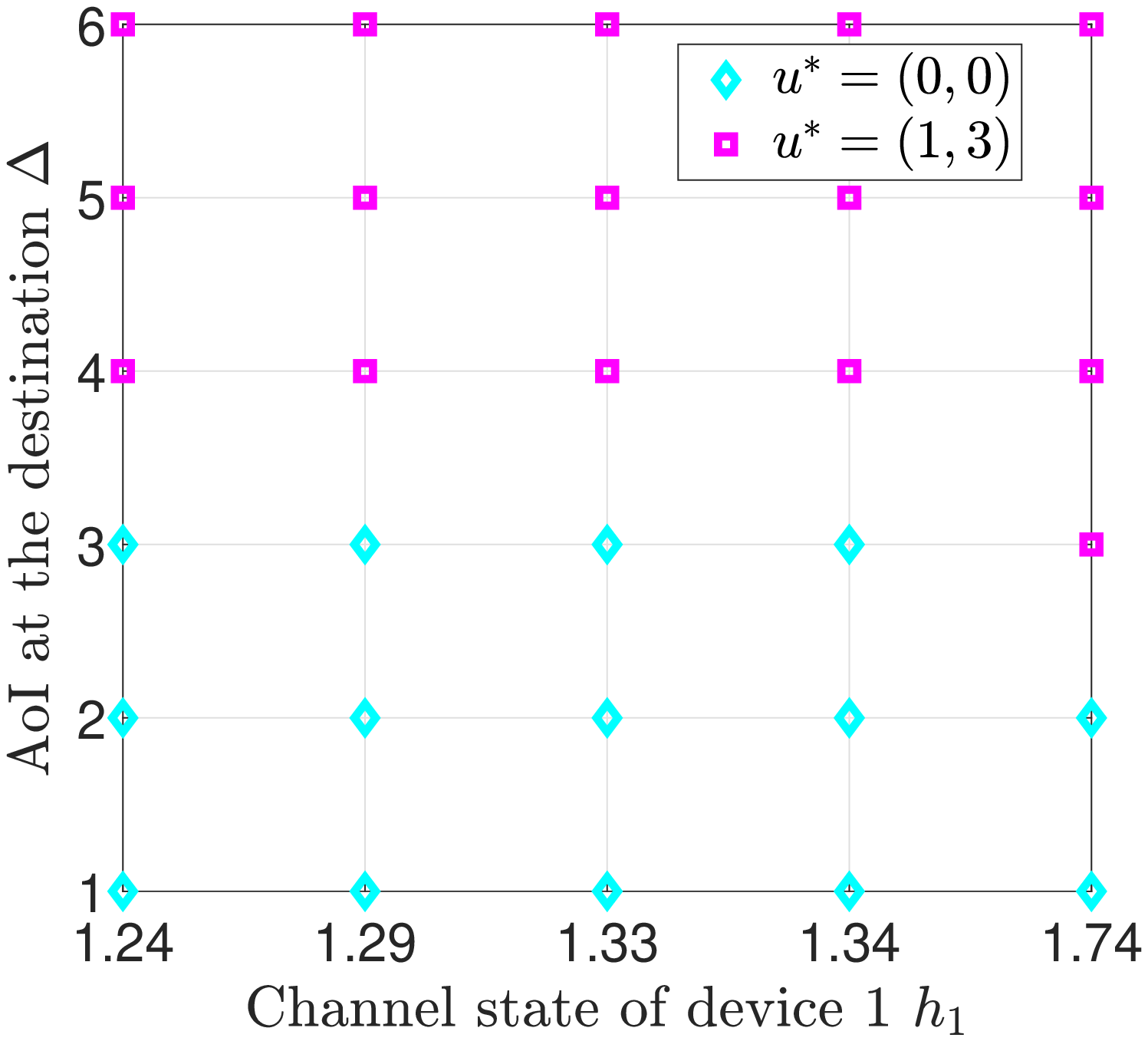}
\subcaption{Fixed $\bs{A}$ and $\bs{h}\setminus h_1$.}\label{fig:structure-h3}
\end{minipage}
\caption{Structure of the optimal scheduling policy $\pi^*$. $N_1=2$, $N_2=3$, $M=2$, $\hat{\Delta}=\hat{A}_n=6$, $|\mathcal{H}_n|=5$, and $\beta_n=1$, $\forall n\in\mathcal{N}$.}\label{fig:general}
\end{figure}

From Theorem~\ref{theorem:optimal}, we know that, for given $\bs{A}$ and $\bs{h}$, the optimal scheduling policy is threshold-based with respect to the AoI at the destination $\Delta$.
Fig.~\ref{fig:general} illustrates the analytical results of Theorem~\ref{theorem:optimal}.
Fig.~\ref{fig:general} shows that, when $\Delta$ is small, it is not efficient to schedule the IoT devices to update their status packets to the destination, as a higher energy cost per age reduction is consumed. Meanwhile, whenever $\Delta$ is large, it is more efficient to schedule the devices, as the previous re-constructed status of the physical process becomes rather outdated.
Moreover, the monotonicity of the threshold $\phi_{\bs{u}^*}(\bs{A},\bs{h})$ in Fig.~\ref{fig:structure-h3} indicates that, when the channel condition is better, it is more efficient to schedule the IoT devices, as a lower energy cost is consumed. Such a threshold-based structure can be exploited to design low-complexity structure-aware optimal algorithms, by following approaches similar to those in \cite{8778671} and \cite{8938128}.

\section{Special Case: IoT Monitoring System with Only Type-II devices}
We now consider a special IoT monitoring system with only type-II devices, i.e., $\mathcal{N}_2=\mathcal{N}$.
Note that, here, the system states consist of only the AoI at the destination $\Delta$ and the system channel state $\bs{h}$.
The dynamics of the AoI at the destination are given by
\begin{align}\label{eqn:aoi_destination_special}
 \Delta(t+1)=\begin{cases}
 1, &\text{if}~\sum_{n\in\mathcal{N}}\bs{u}(t)=M,\\
 \min\{\Delta(t)+1,\hat{\Delta}\},  &\text{otherwise}.                
      \end{cases}
\end{align}
Given an observed  $\Delta$ and $\bs{h}$, the system scheduling action $\bs{u}$ is determined according to a feasible stationary scheduling policy $\pi$ (with some abuse of notation), which is a mapping from the system state space $\varDelta\times\mathcal{H}$ to the feasible system action space $\mathcal{U}$, i.e., $\pi(\Delta,\bs{h})=\bs{u}$.
Under the dynamics \eqref{eqn:aoi_destination_special}, the induced random process $\{\Delta(t),\bs{h}(t)\}$ for a given policy $\pi$ is a controlled Markov chain with the  transition probability $\Pr[\Delta',\bs{h}'|\Delta,\bs{h},\bs{u}]=\Pr[\Delta'|\Delta,\bs{u}]\prod_{n\in\mathcal{N}}\Pr[h'_n]$,
where
\begin{align}\label{eqn:delta_prob}
&\Pr[\Delta'|\Delta,\bs{u}]=\Pr[\Delta'(t+1)=\Delta'|\Delta(t)=\Delta,\bs{u}(t)=\bs{u}]\nonumber\\
&=\begin{cases}
1,  &~\text{if}~ \Delta'\text{satisfies~the~dynamics of~in~\eqref{eqn:aoi_destination_special}},\\
                0, &~\text{otherwise.}
      \end{cases}
\end{align}

As previously done, we can formulate the problem as in \eqref{eqn:mdp} and obtain the associated Bellman equation:
\begin{align}\label{eqn:bellman_special}
&  \theta + V(\Delta,\bs{h}) = \min_{\bs{u}\in\mathcal{U}} \Big\{\Delta + \sum_{n\in\mathcal{N}}\beta_nu_n (C_n^s+C^u_n(h_n))  \nonumber\\
&\hspace{20mm} + \sum_{\Delta,\bs{h}'} \Pr[\Delta',\bs{h}'| \Delta,\bs{h},\bs{u}] V(\Delta',\bs{h}')\Big\}.
\end{align}
The optimal policy $\pi^*$ can be determined by:
\begin{align}\label{eqn:optimal_policy_special}
&  \pi^*(\Delta,\bs{h})= \arg\min_{\bs{u}\in\mathcal{U}} \Big\{\Delta + \sum_{n\in\mathcal{N}}\beta_nu_n (C_n^s+C^u_n(h_n))  \nonumber\\
&\hspace{20mm} + \sum_{\Delta,\bs{h}'} \Pr[\Delta',\bs{h}'| \Delta,\bs{h},\bs{u}] V(\Delta',\bs{h}')\Big\}.
\end{align}

From \eqref{eqn:aoi_destination_special}, we can see that,  for given $(\Delta,\bs{h})$,  the next AoI state at the destination $\Delta'$ is the same for any $M$ scheduled devices. Thus, by \eqref{eqn:optimal_policy_special}, we know that if $\pi^*(\Delta,\bs{h})\neq \bs{0}_N$, then $\pi^*(\Delta,\bs{h})=\bs{u}^{\dag}$ must satisfy that:
\begin{align}
u_n^{\dag}=\begin{cases}
1,  &~\text{if}~ n\in\mathcal{N}(\bs{h}),\\
                0, &~\text{otherwise.}
      \end{cases}
\end{align}
Here,  $\mathcal{N}(\bs{h})= \{i\in\mathcal{N}: |j\in\mathcal{N}: \beta_i(C_i^s + C_i^u(h_i))>\beta_j(C_j^s + C_j^u(h_j))|<M\}$ is the set of the devices having the $M$ smallest values of the weighted energy cost $\beta_n (C^s_n+C^u_n(h_n))$.
Now, for each $\bs{h}\in\mathcal{H}$, define $C_h\triangleq\sum_{n\in\mathcal{N}(\bs{h})}\beta_n(C^s_n+C^u_n(h_n))$ as the energy cost state of all IoT devices. Let $\mathcal{C}_h$ be the set of all possible $C_h$.
Clearly,  the optimal control policy $\pi^*$ in \eqref{eqn:optimal_policy_special} depends on $\Delta$ and $C_h$.
Thus, we can obtain an equivalent MDP, in which, the system state space is $\varDelta\times\mathcal{C}_h$ and the action space is $\tilde{\mathcal{U}}(C_h)=\{\bs{0}_N,\bs{u}^{\dag}\}$.
It can be seen that the equivalent MDP has smaller  state and action spaces as  $|\mathcal{C}_h|\leq |\mathcal{H}|$ and $|\tilde{\mathcal{U}}(C_h)|=2 \leq |\mathcal{U}|=1+ \binom{N}{M}$.
Then, we have the corresponding Bellman equation as follows:
\begin{align}\label{eqn:bellman_special_reduced}
  \theta + V(\Delta,C_h) = \min \Big\{&\Delta +  \mathbb{E}[V(\min\{\Delta+1,\hat{\Delta}\},C_h')],\nonumber\\
&\Delta + C_h + \mathbb{E}[V(1,C_h')]\Big\},
\end{align}
where the expectation is taken over the distribution of $C_h$.

Along the structural analysis in Section~\ref{sec:structure}, we characterize the structural properties of the optimal policy $\pi^*$ for the special case with only type-II devices.
\begin{theorem}\label{theorem:optimal_special}
For each $(\Delta,C_h)\in\varDelta\times\mathcal{C}_h$, there exists a threshold $\psi(C_h)$ such that
\begin{align}\label{eqn:structure_threshold}
\pi^*(\Delta,C_h)=\begin{cases} \bs{u}^{\dag},  & \text{if}~\Delta\geq \psi(C_h), \\
            \bs{0}_N,  &\text{otherwise}.
  \end{cases}
\end{align}
In \eqref{eqn:structure_threshold}, $\psi(C_h)$ is non-decreasing with $C_h$.
\end{theorem}

\begin{figure}[!t]
\begin{centering}
\includegraphics[scale=.4]{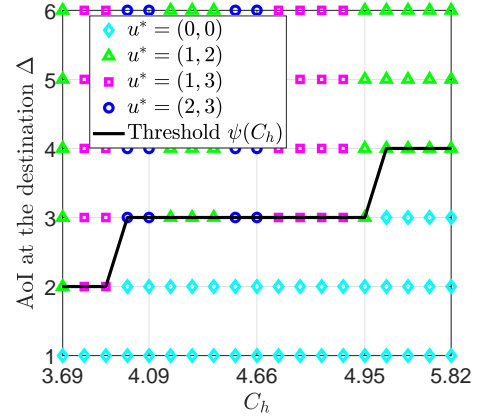}
 \caption{Structure of the optimal scheduling policy $\pi^*$ for the special case of an IoT with only type-II devices. $N=3$, $M=2$, $\hat{\Delta}=\hat{A}_n=6$, $|\mathcal{H}_n|=4$, and $\beta_n=1$, $\forall n$.}\label{fig:special}
\end{centering}
\end{figure}

Theorem~\ref{theorem:optimal_special} reveals that the scheduling action is threshold-based with respect to $\Delta$, as illustrated in Fig.~\ref{fig:special}.
Moreover, the monotonicity property of $\psi(C_h)$ shown in Fig.~\ref{fig:special} indicates that it is more efficient to schedule the IoT devices when the channel condition is better, as $C_h$ is non-increasing with $\bs{h}$.

\section{Simulation Results and Analysis}
In this section, we compare the performance of the optimal policy with a myopic baseline policy for the system in which there are $N_1=2$ type-I devices and $N_2=3$ type-II devices, and  the status packets from $M=2$ devices are needed at the destination.
For the myopic policy, in each slot, we choose the scheduling action that maximizes the AoI reduction at the destination minus the weighted energy cost, i.e.,
$\bs{u}(t)= \arg\min_{\bs{u}(t)\in\mathcal{U}} \Delta(t) - \Delta(t+1) - \sum_{n\in\mathcal{N}}\beta_nu_n(t)(C^s_n+C^u_n(h_n(t))) $.
We consider that there are in total 4 possible channel states for each device, which are randomly selected from $[1,2]$ and the corresponding distribution probabilities are $\{1/4,1/4,1/4,1/4\}$.
We assume that the arrival rate $\lambda_n$ for each type-I device  is randomly selected from $[0.3,0.8]$, the updating cost for each device is $C_n^u(h_n)=C_n^u/h_n$, where $C_n^u$ is randomly selected from $[2,3]$, and the sampling cost $C_n^s$ for each type-II device is randomly selected from $[1,2]$.
The weighting factors are assumed to be same for all devices.

\begin{figure*}[!t]
\begin{minipage}[h]{0.33\linewidth}
\centering
       \includegraphics[scale=0.38]{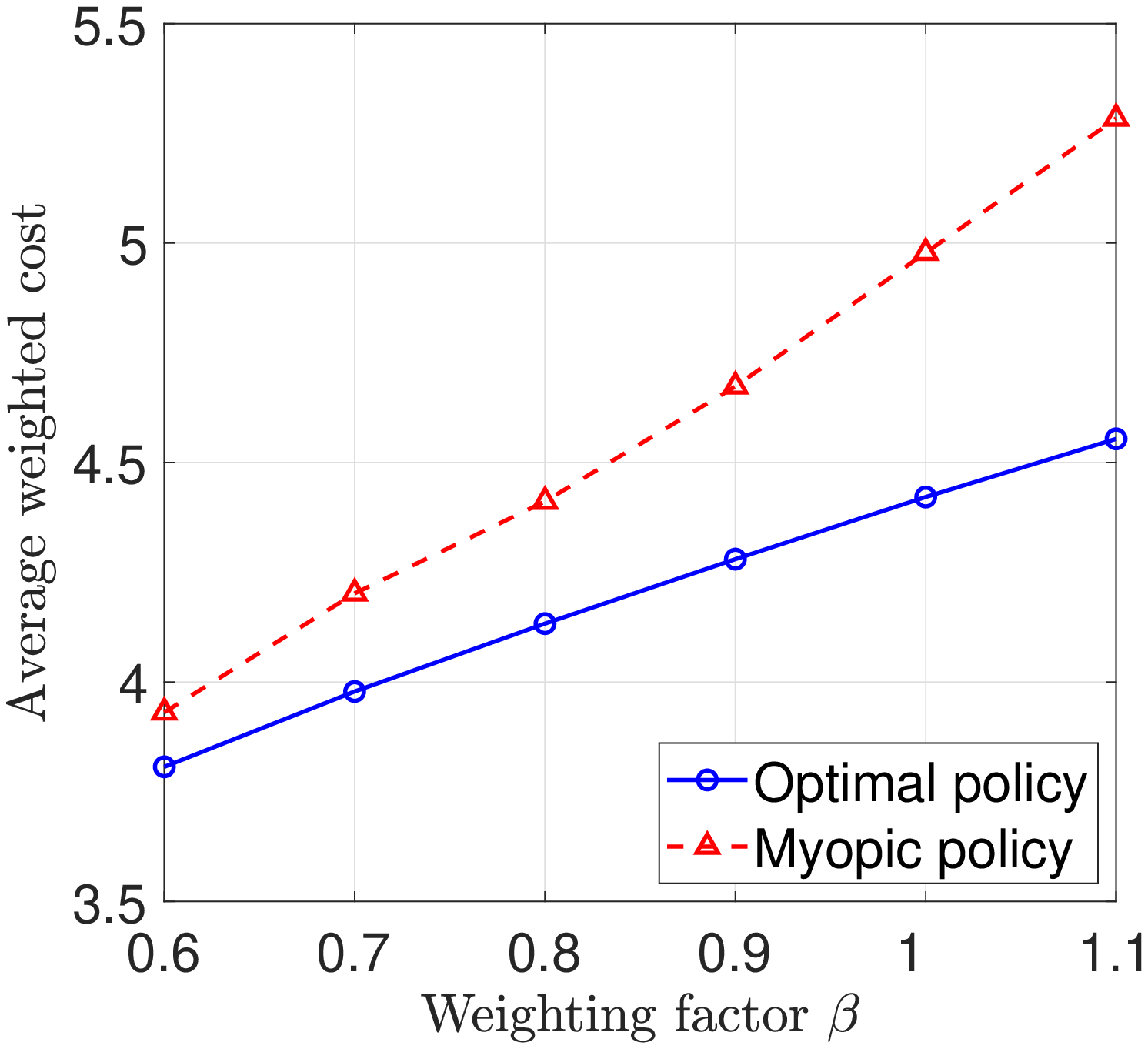}
\subcaption{Average weighted cost.}\label{fig:sum-cost-vs-beta}
\end{minipage}
\begin{minipage}[h]{0.33\linewidth}
\centering
        \includegraphics[scale=0.38]{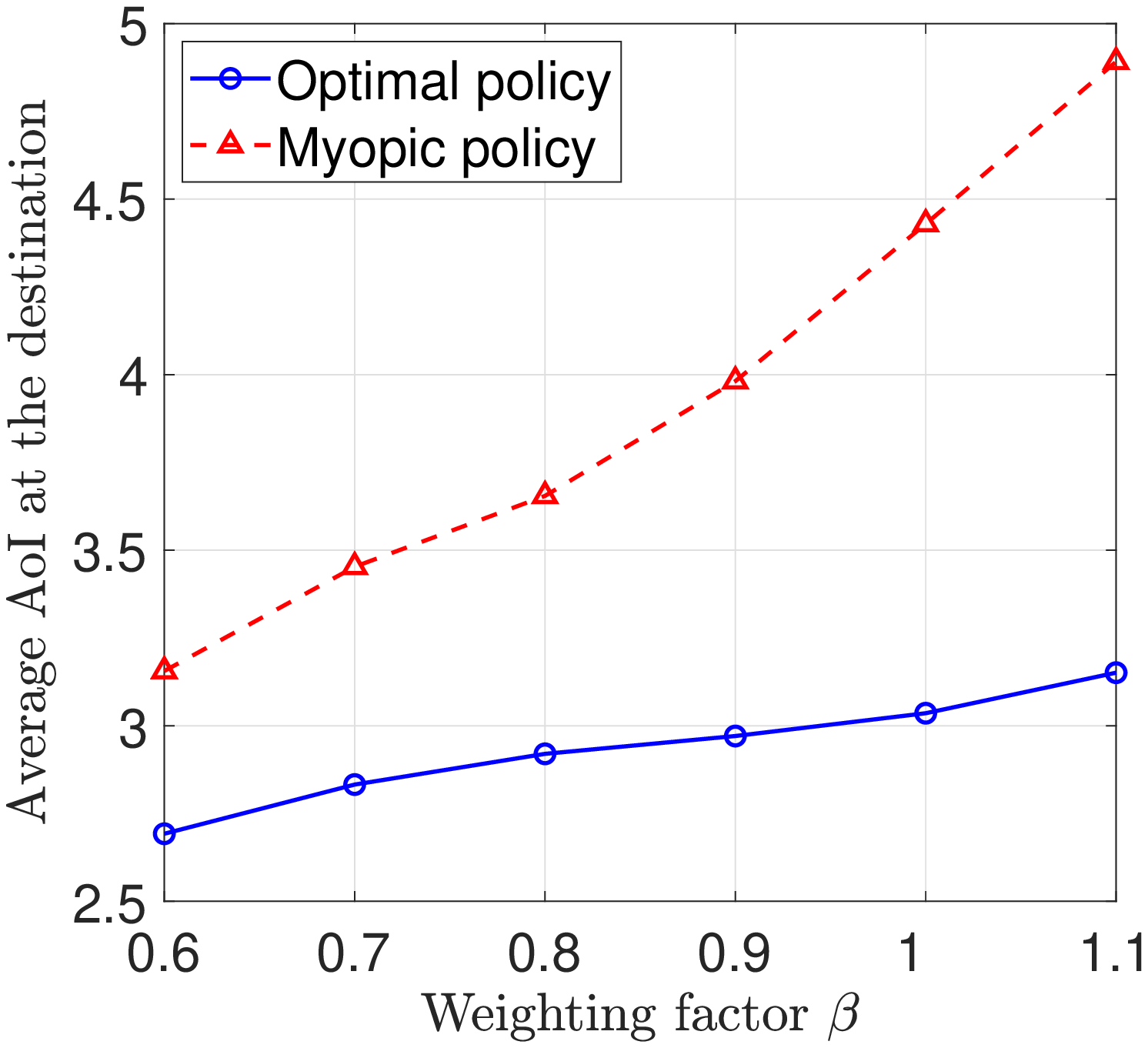}
\subcaption{Average AoI at the destination.}\label{fig:AoI-vs-beta}
\end{minipage}
\begin{minipage}[h]{0.33\linewidth}
\centering
        \includegraphics[scale=0.38]{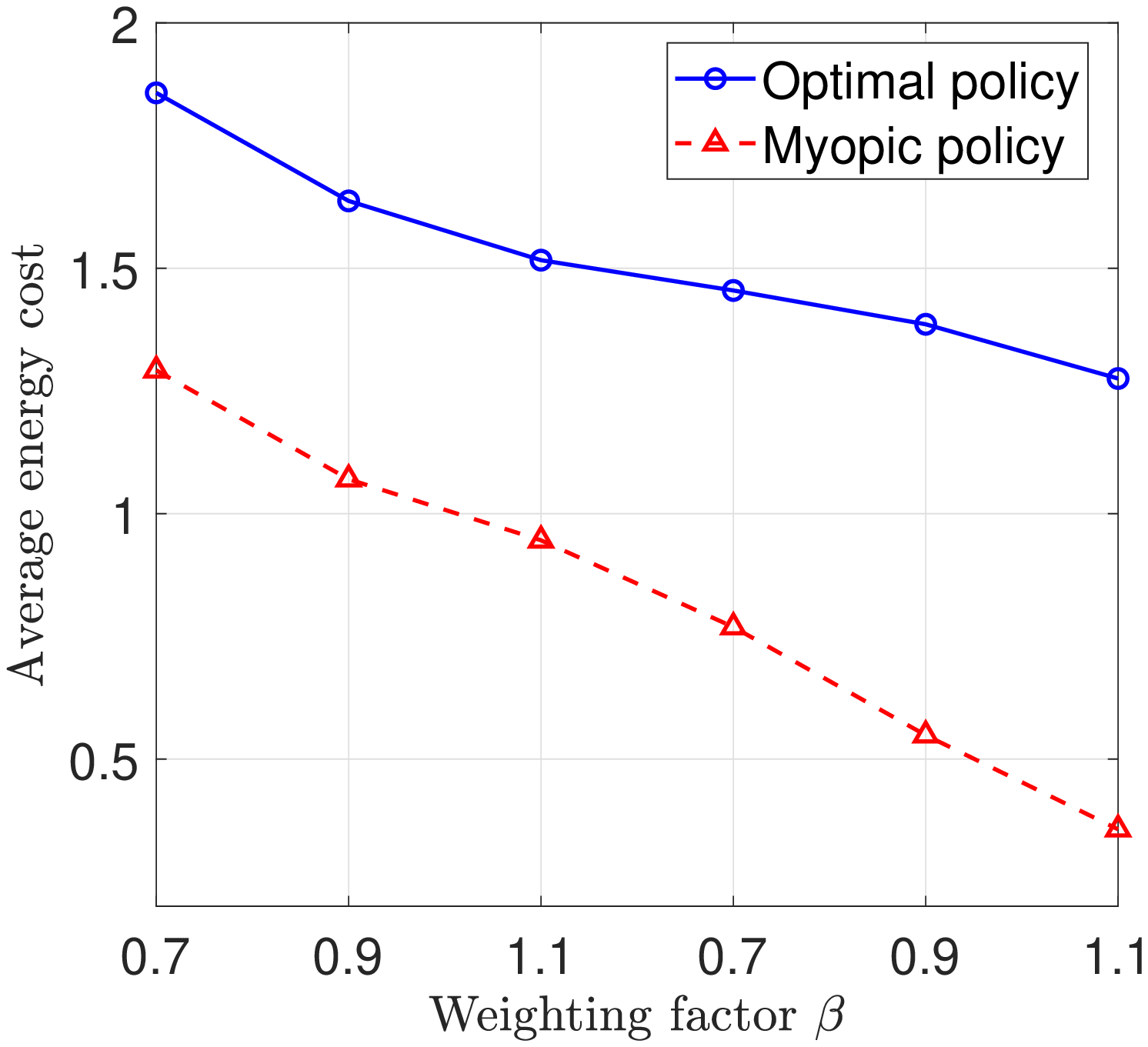}
\subcaption{Average energy cost at the devices.}\label{fig:energy-cost-vs-beta}
\end{minipage}
\caption{Performance comparison between the optimal policy and the myopic baseline policy.}\label{fig:cost-comparison}
\end{figure*}

Fig.~\ref{fig:cost-comparison} shows the average weighted cost, the average AoI at the destination, and the average energy cost, resulting from the optimal policy and the myopic baseline policy, under different weighting factors $\beta$.
The simulation results are obtained by averaging over $50,000$ time slots.
Fig.~\ref{fig:cost-comparison} demonstrates that the optimal policy can respectively reduce the average weighed cost and the average AoI at the destination by up to 14\% and 36\%  over the myopic baseline policy, with a higher energy cost.
Hence, the optimal policy can make better decisions by fully utilizing the energy at the IoT device.
Moreover, for both polices, when the weighting factor $\beta$ increases, the average AoI at the destination increases while the average energy cost decreases. This reveals the tradeoff between the AoI at the destination and the energy cost at the devices.

\section{Conclusion}
In this paper, we have studied the optimal device scheduling process that jointly minimizes the average AoI at the destination and the energy cost at the devices for a real-time IoT monitoring system with two types of correlated devices.
We have formulated this problem as an infinite horizon MDP and shown that the optimal policy is threshold-based with respect to the AoI at the destination and the threshold is non-increasing with the channel condition.
For a special case with only type-II devices, we have reduced the original MDP to an MDP with smaller state and action spaces, and shown a threshold-based structure of the optimal policy, for which, the threshold is non-decreasing with an energy cost function.
Simulation results have verified the analytical results and illustrated the effectiveness of the optimal policy.

\bibliographystyle{IEEEtran}
\bibliography{IEEEabrv,ref}

\end{document}